\documentclass[]{spie}  

\usepackage{amsmath,amsfonts,amssymb}
\usepackage{graphicx}
\usepackage[colorlinks=true, allcolors=blue]{hyperref}
\usepackage{subcaption}

\title{A Near-Infrared Pyramid Wavefront Sensor for the MMT}

\author[a,b]{Jacob Taylor}

\author[a,b]{Suresh Sivanandam}
\author[c]{Narsireddy Anugu}
\author[a,b]{Adam Butko}
\author[b]{Shaojie Chen}
\author[d]{Olivier Durney}
\author[f]{Tim Hardy}
\author[b]{Masen Lamb}
\author[d]{Manny Montoya}
\author[d]{Katie Morzinski}
\author[b,e]{Robin Swanson}

\affil[a]{David A. Dunlap Department of Astronomy \& Astrophysics, University of Toronto}
\affil[b]{Dunlap Institute for Astronomy \& Astrophysics}
\affil[c]{Center for High Angular Resolution Astronomy}
\affil[d]{Steward Observatory, University of Arizona}
\affil[e]{Department of Computer Science, University of Toronto}
\affil[f]{NRC Herzberg Astronomy and Astrophysics Research Centre}

\authorinfo{Further author information: (Send correspondence to J.T.)\\J.T.: E-mail: jacob.taylor@mail.utoronto.ca}

\begin{document} 
\maketitle

\begin{abstract}
The MMTO Adaptive optics exoPlanet characterization System (MAPS) is an ongoing upgrade to the 6.5-meter MMT Observatory on Mount Hopkins in Arizona. MAPS includes  an upgraded adaptive secondary mirror (ASM), upgrades to the ARIES spectrograph, and a new AO system containing both an optical and near-infrared (NIR; 0.9-1.8 µm) pyramid wavefront sensor (PyWFS). The NIR PyWFS will utilize an IR-optimized double pyramid coupled with a SAPHIRA detector: a low-read noise electron Avalanche Photodiode (eAPD) array. This NIR PyWFS will improve MAPS’s sky coverage by an order of magnitude by allowing redder guide stars (e.g. K \& M-dwarfs or highly obscured stars in the Galactic plane) to be used.  To date, the custom designed cryogenic SAPHIRA camera has been fully characterized and can reach sub-electron read noise at high avalanche gain. In order to test the performance of the camera in a closed-loop environment prior to delivery to the observatory, an AO testbed was designed and constructed.  In addition to testing the SAPHIRA’s performance, the testbed will be used to test and further develop the proposed on-sky calibration procedure for MMTO’s ASM. We will report on the anticipated performance improvements from our NIR PyWFS, the SAPHIRA's closed-loop performance on our testbed, and the status of our ASM calibration procedure.
\end{abstract}


\section{INTRODUCTION} \label{sec:intro}

As pristine starlight passes through the atmosphere, its wavefront becomes undulated and distorted which ultimately limits the resolution of large ground-based telescopes.
Adaptive optics (AO) is a method of improving the quality of ground-based astronomical observations by rapidly measuring and correcting the wavefront error introduced by turbulence in the Earth's atmosphere.
In recent decades, AO systems have become more sensitive, robust, and prevalent; moreover, the next-generation of extremely large telescopes will be reliant on AO to achieve many of their key scientific goals.
While there is substantial variety across AO systems, they all have three key components: a wavefront sensor (WFS), a deformable mirror (DM), and a real-time controller (RTC).
These components typically work together in a `closed-loop' configuration to compute and correct atmospheric distortions in real-time.thesis 

The MMT observatory, situated on Mount Hopkins in Arizona, hosts a 6.5m primary mirror and has had an operational AO system since 2002 \cite{MMTO_ASM_old}; however, the system has reached the end of its lifetime and a new system has been commissioned. 
The upgraded AO system is part of the MMTO Adaptive optics exoPlanet characterization System (MAPS) which consists of a new AO system, a refurbished high-resolution IR spectrograph (ARIES, 1-5$\mu$m) and a new AO-optimized upgrade to the MMT-sensitive polarimeter (MMT-Pol) \cite{MAPS_status}.
The project's main science goal is to characterize exoplanet atmospheres through the marriage of a state-of-the-art AO system and broad wavelength, high-resolution IR spectroscopy (HRS).
By cross-correlating spectra with molecular templates, MAPS will be able to characterize  atmospheric properties of exoplanets.
Please see proceedings by Montoya, M et al. at this conference for more details on the project as a whole.
MMT's new AO system comes with some significant upgrades over its predecessor.
All 336 voice coil actuators of the MMT adaptive secondary mirror (ASM) have been replaced with a new design of actuators and control electronics,
and the legacy 12x12 Shack-Hartmann wavefront sensor has been replaced by two high-sensitivity pyramid wavefront sensors (one optimized for visible wavelengths and the other infrared).
The near-infrared pyramid wavefront sensor (NIR PyWFS) is of particular importance for MAPS since it will drastically increase the instrument's sky coverage. 

The atmosphere's rapid time-evolution, combined with the faintness of most stars, limits natural guide-star (NGS) AO systems to the fraction of the sky containing sufficiently bright sources whose wavefront can be easily measured.
For this reason, while NGS adaptive optics continues to be a powerful tool for astronomers, its applicability is not universal;
however, recently there have been some key advancements in infrared detectors which have the potential to significantly increase NGS AO sky-coverage.
Astronomers now have access to high-speed IR electron avalanche photodiode (eAPD) arrays that can reach sub-electron read noise (e.g. Leonardo's SAPHIRA  \cite{SAPHIRA}).
eAPD arrays are able to reduce the CMOS read noise barrier of near infrared sensors by applying a reverse-bias voltage across each pixel’s P-N junction. 
This voltage accelerates photoelectrons along the electric field, knocking off additional electrons.
By increasing the number of electrons prior to readout, the eAPD arrays are able to achieve sub-electron effective read noise at millisecond readout rates.
These detectors have paved the way for fast and efficient wavefront sensing in the near-infrared which, in turn, significantly improves sky coverage by utilizing guide stars which are normally obscured behind interstellar dust as well as an abundance of red stars (type K and M) whose brightness peaks in the infrared \cite{NIR_WFS}.
The MAPS NIR PyWFS will operate predominately in the J-band, operate at speeds up to 1Khz, and is anticipated to use guide stars down to 18th magnitude (R-band).
This paper will provide an overview of the MAPS NIR PyWFS and an update on it's current commissioning status.

\section{INSTRUMENTATION}

\begin{figure} [ht]
\begin{center}
\begin{tabular}{c} 
\includegraphics[height=10cm]{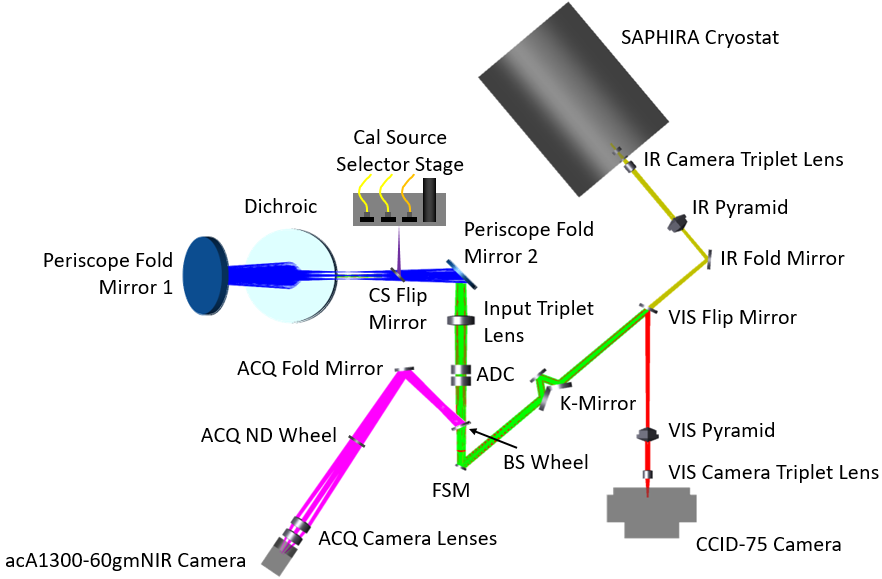}
\end{tabular}
\end{center}
\caption[example] 
{A labelled diagram of the MAPS WFS optical design. The design is divided into 5 channels: periscope channel (blue); baseline channel (green); acquisition channel (purple); visible PyWFS channel (red); NIR PyWFS channel (yellow). The majority of the MAPS WFS unit has been fabricated and installed and the system is currently undergoing optical alignment.\label{fig:maps_optics} }
\end{figure}

The MAPS AO system consists of three main components: a 336 voice coil actuator adaptive secondary mirror; a visible pyramid wavefront sensor ($\sim$ 0.6-0.8 um); and a near infrared pyramid wavefront sensor ($\sim$ 0.9-1.8 um). 
Both WFSs are housed together in the MAPS WFS unit with a flip mirror that will select which will be active.
Additionally, both WFSs will be modulated using the same Physik Instrumente (PI) S-330 Tip Tilt Platform and E-727 Digital Multi-Channel Piezo Controller.
Figure~\ref{fig:maps_optics} shows a diagram of the MAPS WFS unit.
See Montoya, M et al. at this conference for updates on the MAPS adaptive secondary and MAPS visible wavefront sensor.

\subsection{SAPHIRA DETECTOR AND CRYOSTAT}

\begin{figure} [ht]
\begin{center}
\begin{tabular}{c} 
\includegraphics[height=9cm]{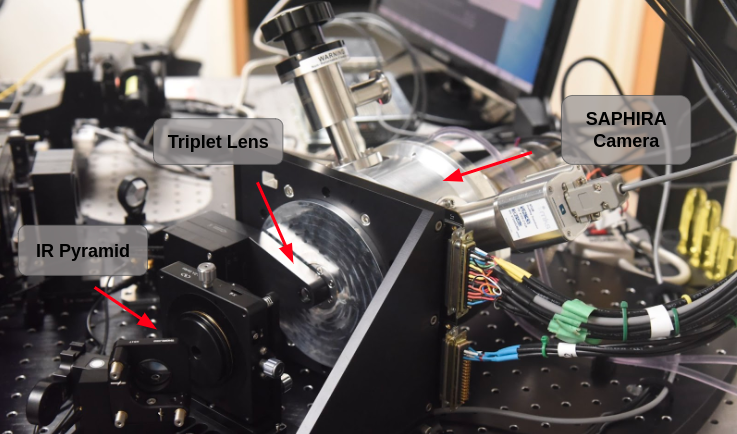}
\end{tabular}
\end{center}
\caption[example] 
{ A photograph of the SAPHIRA cryostat assembly installed into the MAPS WFS unit.\label{fig:maps_wfs_photo} }
\end{figure}

Butko et al. 2018\cite{SHARP_MAPS2} reported on the specifics of our SAPHIRA control system implementation and cryostat chamber which we summarize here.
The chosen detector for the MAPS NIR PyWFS is Leonardo's SAPHIRA Mark 13, a 320 x 256 pixel format HgCdTe CMOS detector integrated with electron avalanche photodiodes (eAPD).
The SAPHIRA is controlled by a custom Astronomical Research Camera (ARC) controller.
The ARC controller utilizes two ARC-46 IR video processing board links to read out the detector in 16 channels (an additional one was installed since Butko et al. 2018\cite{SHARP_MAPS2}).
Both boards are required to achieve the on-sky frame rate goal of 1 kHz using an 80x80 region of interest (ROI) while employing correlated double sampling (CDS) ($\sim 0.48ms$ per read).
Additionally, our controller consists of several other ARC boards: an ARC-64 PCIe board and ARC-22 timing board with three fiber links for fast data transfer; an ARC-32 CCD \& IR clock driver board which provides clock signals and voltage supplies for the detector; and an ARC-33 bias board capable of providing a 0 to 14 V bias for avalanche gain.
The detector is installed on a preamplifier board with differential preamplifier channels, bias decoupling and over-voltage protection which is placed inside a cryogenic Dewar.
The assembly is shown installed inside the MAPS WFS unit in Figure~\ref{fig:maps_wfs_photo}.

The SAPHIRA sensor is housed within a cryostat Dewar, designed by the National Research Council (NRC) Hertzberg, that cools the detector to its operating temperature of 85K.
The cryostat utilizes a Stirling cooler (Sunpower’s CryoTel MT Cryocooler) with an Active Vibration Cancellation (AVC) system.
Butko et al. 2018\cite{SHARP_MAPS2} reports on the Dewar design, cryocooler performance and vibration testing.

\subsection{AO TESTBED}

The Renovated Adaptive Z-band Optic Relay system (RAZOR) is an adaptive optics testbed located at the University of Toronto's Spectroscopic High Angular Resolution/Photonics (SHARP) imaging lab. 
RAZOR was constructed to test novel AO techniques and technologies in a controlled environment prior to their deployment at an observatory; in particular, the closed-loop performance testing of the SAPHIRA sensor.
RAZOR has been designed to test emerging techniques and technologies from various AO sub-fields: the testbed hosts three independent wavefront sensing arms; operates at both optical and near-infrared wavelengths; and employs a custom real-time control software developed in house.
These specifications provide the flexibility and capability required to test and develop a range of cutting edge AO techniques.

\begin{figure} [ht]
\begin{center}
\begin{tabular}{c} 
\includegraphics[width=0.9\textwidth]{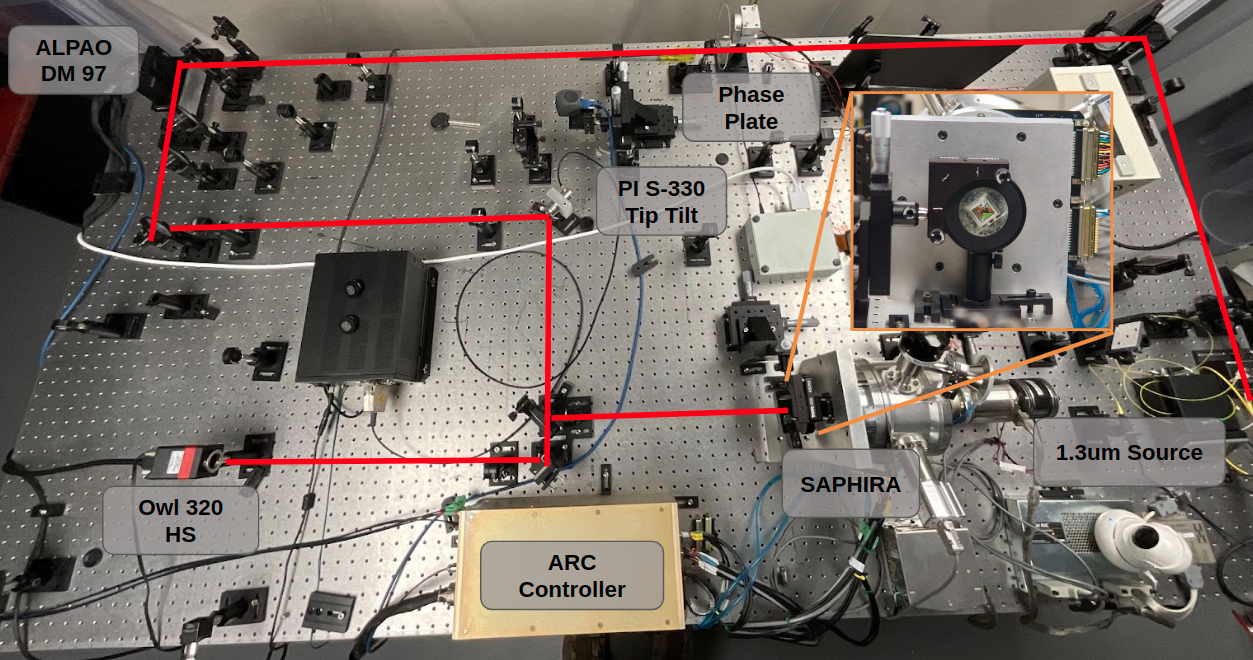}
\end{tabular}
\end{center}
\caption[example] 
{ \label{fig:razor} A photo of our adaptive optics testbed (RAZOR) from above. The light path for the experiments reported in this proceedings is shown in red and the relevant components are labelled. The orange box shows a zoomed in view of the pyramid.}
\end{figure} 

RAZOR uses two co-aligned monochromatic point sources to facilitate operation in both the visible and near-infrared: one 650nm red laser and a 1.31um infrared laser that can be swapped for an 890nm superluminescent diode.
The testbed makes use of achromatic lenses so that both wavelengths can be used simultaneously with minimal chromatic error. 
Both sources pass through simulated atmosphere --- one of four possible rotating phase screens with $r_0$ between 0.5 and 1.5 millimeters --- before illuminating the surface of a 97 actuator ALPAO deformable mirror (DM97-15).
After reflecting off of the DM97, the light is separated into three independent wavefront sensing arms: a HASO4 First Shack-Hartmann wavefront sensor (SHWFS) from Imagine Optic, a custom-built SHWFS arm, and a PyWFS arm.
The PyWFS arm is composed of a PBL25Y/S-BAM12 glass Pyramid (spare copy of MAPS pyramid with manufacturing defects), a tip-tilt modulator (PI S-330 Tip Tilt and E-727 Controller), and the SAPHIRA camera. 
To quantify the quality of the correction, RAZOR's `science' arm utilizes a XIMEA CMOS camera (MQ013MG-ON) or a Raptor Photonics InGaAs CMOS camera (Owl 320 HS) depending on the operating wavelength.

RAZOR is controlled through an in-house real-time control software called \textit{SharpRTC}. 
The core components of SharpRTC are written in \textit{C/C++} with an emphasis on object-oriented design. 
Each hardware component is implemented as a stand alone `object class' and interactions between hardware components are implemented as a `process class'.
At runtime, each object class is instantiated with parameters extracted from a user defined configuration file (YAML format). 
The configuration file also determines which calibration or pre-processing features will be executed before running the closed AO loop.
This design gives the user a large amount of customizability surrounding the many parameters of an AO loop as well as providing a clear path for adding additional functionality.
In addition to the core components, SharpRTC also includes a variety of analysis scripts written in \textit{Python}. 
The analysis scripts contain tools for probing performance parameters such as loop latency, slope diagnostics, modulation shape, and Strehl-ratios. 
Together, all of the SharpRTC components offer the user a significant amount of flexibility in their operation and caters to a wide variety of AO testing and development.

\section{CHARACTERIZATION \& RESULTS}

\subsection{SAPHIRA Characterization}

In early 2020, a suite of characterization tests were performed on our SAPHIRA detector in order to determine each pixel's avalanche gain, conversion gain, nominal read noise, effective read noise, and excess noise.
Since we only require a small region of interest on the detector for wavefront sensing, we purchased an engineering grade SAPHIRA that is known to have regions with sub-optimal performance; therefore, the purpose of the characterization was to verify performance and to determine the optimal region of interest.

\begin{figure} [ht]
\begin{center}
\begin{tabular}{c} 
\includegraphics[height=7cm]{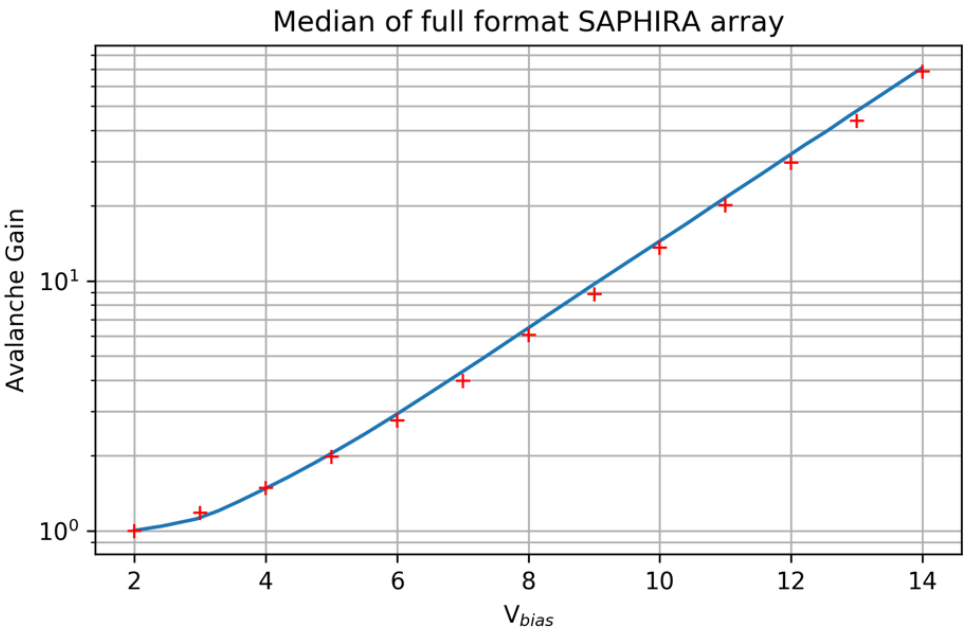}
\end{tabular}
\end{center}
\caption[example] 
{Median avalanche gain across the array as function of reverse-bias voltage. The blue curve represents the reported avalanche gain for SAPHIRA Mark 3 \& 10 arrays. \label{fig:avalanche}}
\end{figure}

The first characterization parameter of interest is avalanche gain ($M$) --- the multiplicative factor of photoelectrons resulting from impact ionization --- as a function of the reverse-bias voltage, $V_{bias}$.
Light frames were acquired using a near-infrared LED; 128 exposures were acquired with 100 Samples Up the Ramp (SUTR) each.
To account for dark current, background frames were captured using the same exposure time as light frames but with the NIR LED off.
The second read from each exposure was used to perform CDS and calibrate out each pixel’s pedestal voltage.
The avalanche gain for each pixel was obtained by dividing the linear slope (in ADU/s) by the calibrated slope that yields unity avalanche gain ($V_{bias} = 2V$). 
The results agree with previous studies and are summarized in Figure~\ref{fig:avalanche}.

\begin{figure} [ht]
\begin{center}
\begin{tabular}{c} 
\includegraphics[height=7cm]{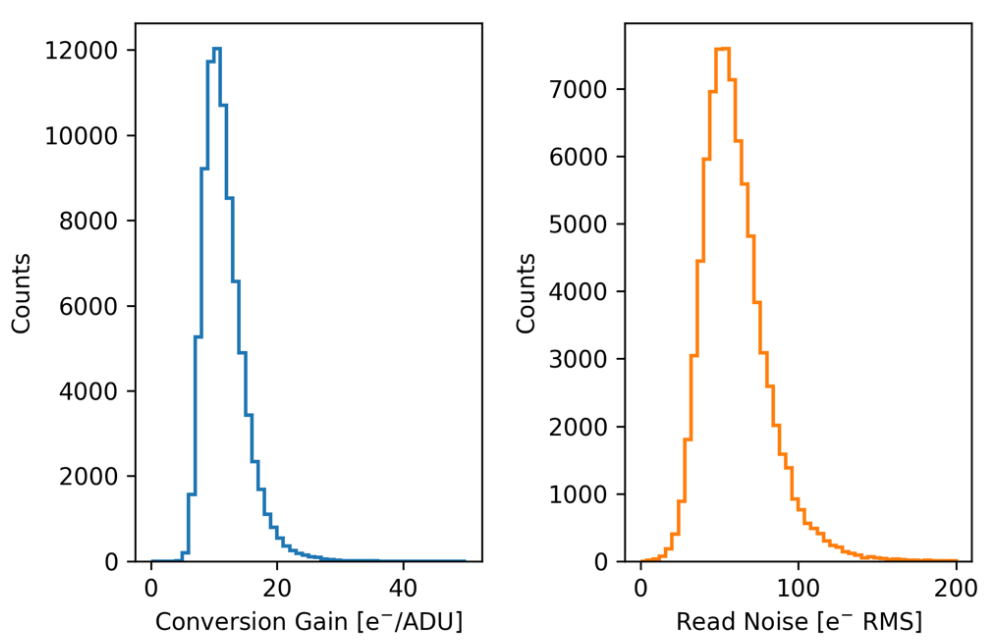}
\end{tabular}
\end{center}
\caption[example] 
{Histograms showing each pixel's measured conversion gain (left) and read noise (right) for our SAPHIRA detector. The median conversion gain and read noise of the detector were found to be $11 \frac{e^-}{ADU}$ and $58 e^-$ respectively.\label{fig:readnoise_gain} }
\end{figure}

Conversion gain and read noise for the array were determined using the Photon Transfer Curve (PTC) method.
The same data used to calculate the avalanche gain were also used to compute the conversion gain.
Each pixel's PTC was estimated by plotting the pixel variance as a function of mean pixel value across the 128 exposures.
A line was fit to the linear regime of the PTC and the conversion gain was taken to be the inverse slope of the fit.
To measure the read noise, a similar process was conducted using dark frames to better populate the read noise dominated regime.
The dark frames were acquired with a cooled aluminum blank filter installed in front of the array to reduce the amount of NIR background light illuminating the detector.
From the resulting PTC, the y-intercept of the linear-fit was converted to electrons RMS using the pixel’s conversion gain (an additional factor of $\sqrt{2}$ is needed to account for the CDS).
The median conversion gain of the detector was found to be $11 \frac{e^-}{ADU}$ and the median read noise was found to be $58 e^-$.
The full distributions of both parameters are shown in Figure~\ref{fig:readnoise_gain}.

Using the nominal read noise, conversion gain, and avalanche gain we are able to estimate the effective read noise using the following relation:

\begin{equation}
    ERN_{e^-} (V_{bias}) = G\frac{\sigma (V_{bias}) }{M (V_{bias})}
\end{equation}

Where $G$ is the pixels conversion gain, $\sigma$ is the pixel's read noise in ADU, and $M$ is the pixel's avalanche gain. We found that a median effective read noise of $0.85 e^-$ is achieved with a bias voltage of 14V.

\begin{figure} [ht]
\begin{center}
\begin{tabular}{c} 
\includegraphics[height=7cm]{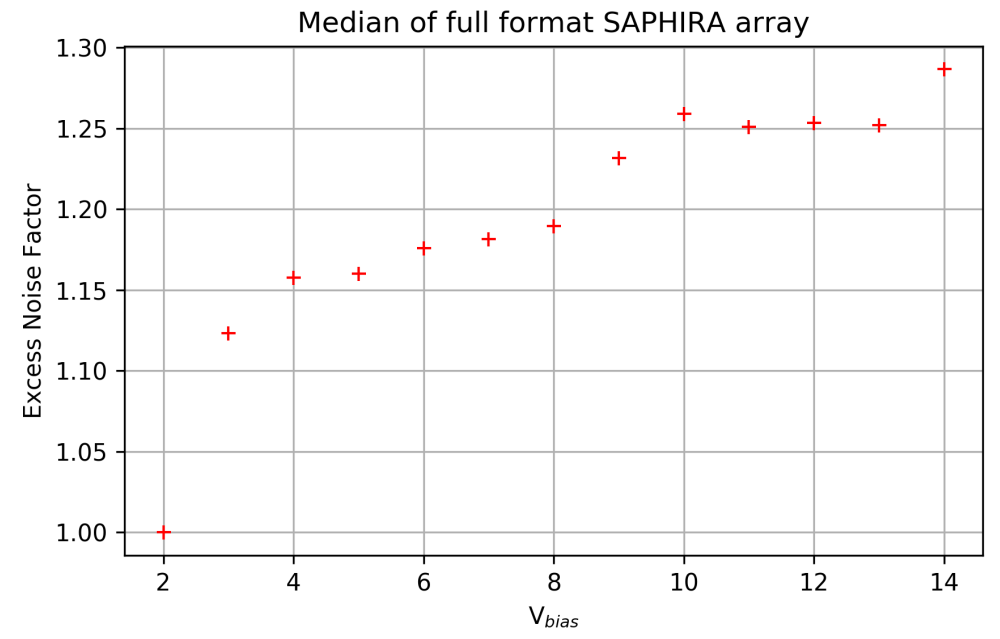}
\end{tabular}
\end{center}
\caption[example] 
{The median excess noise factor across the array as a function of applied bias voltage. There is a more scatter in the excess noise measurement after a bias voltage of 8V. It is possible this is a result of having fewer points within the Poisson-dominated regime of the PTCs for these trials, and therefore, less accuracy in our slope measurements. \label{fig:excess_noise} }
\end{figure}

The impact ionization responsible for avalanche gain is a stochastic process that introduces some excess noise. 
This excess noise degrades the signal-to-noise ratio and can be quantified as the ratio, $F$, of the observed photon noise to what we would expect if there was no excess noise.
The excess noise can be measured as the quotient of the slopes of PTCs in the Poisson-noise regime for $V_{bias} > 2V$ and the product of each pixel’s conversion and avalanche gains for that bias.
The median excess noise factors across the array are summarized in Figure~\ref{fig:excess_noise}.

\subsection{Closed-loop Performance}

Prior to shipping the SAPHIRA and cryostat assembly to be integrated into the MAPS wavefront sensor unit at the University of Arizona, we tested the camera in closed-loop on our AO testbed.
The camera was illuminated using the 1.31um laser source and the PSF was monitored using the Owl 320 HS infrared camera from Raptor Photonics.
In order to best replicate MMT atmospheric conditions, the tests were conducted mimicking our expected aperture size to atmosphere coherence length ratio at the MMT ($\frac{D}{r_0} \approx 13.5$), and phase plate rotation speeds were chosen to be analogous to typical wind speeds ($10mm/s$ on the phase plate $\approx 10m/s$ at MMTO).
We were able to test for minimum required functionality and did not test the PyWFS with modulation.
Additional testing will be performed once the camera is fully integrated into the MAPS WFS unit.

Our tests employed a standard AO control loop.
An empirical interaction matrix (IM) was computed using the poke/pull method whereby each actuator is repeatedly perturbed and the slope response is recorded.
The command matrix (CM) was computed via a Singular Value Decomposition (SVD) of the IM.
To improve stability, up to 30 modes were dropped during the SVD; we believe the lack of modulation and imperfections on the phase screen were the main causes of the instabilities, both of which should not be a problem under nominal operation at MMTO.
The loop was run 1.1Khz, limited by the readout speed of the detector (ROI of 96x96 pixels with dark subtraction and deinterlacing).

\begin{figure} [ht]
\begin{center}
\begin{tabular}{c} 
\includegraphics[height=7cm]{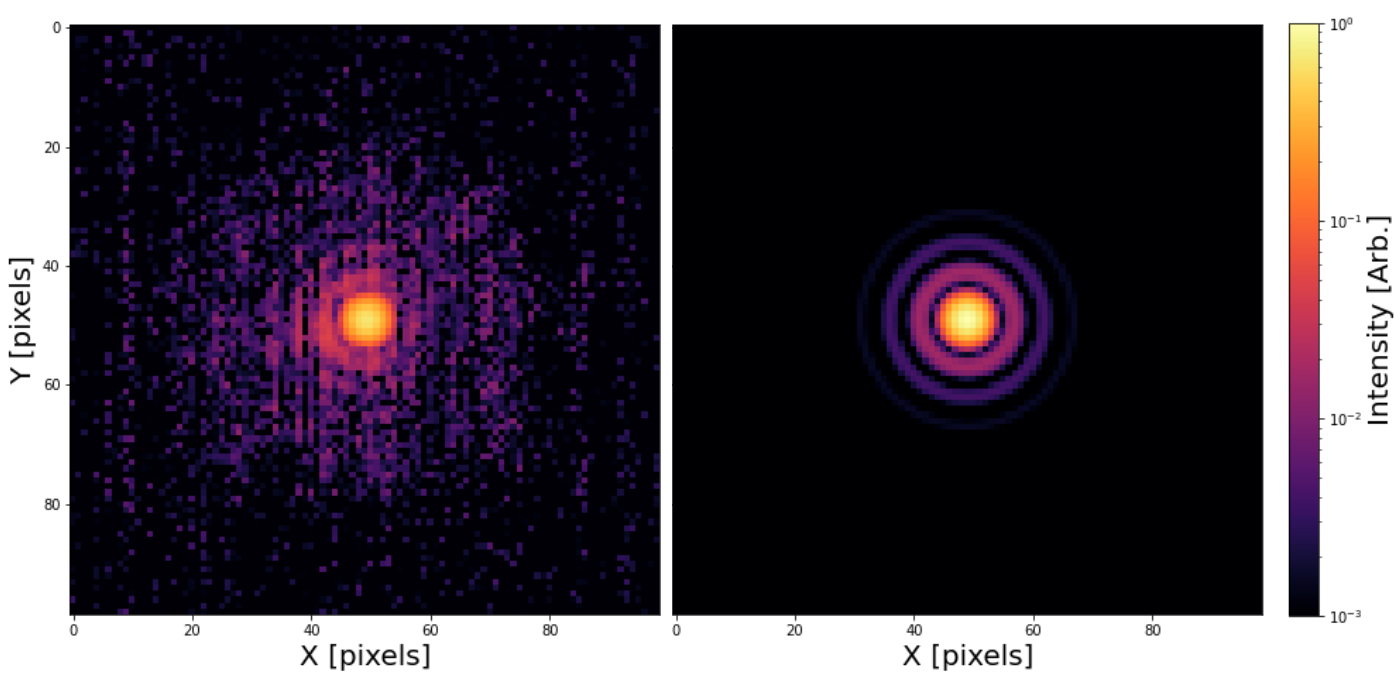}
\end{tabular}
\end{center}
\caption[example] 
{A comparison between the observed closed loop PSF (1000 stacked PSF frames taken over one phase screen rotation, estimate Strehl $\approx$ 65\%) and the down-sampled best-fit airy model for the underrated testbed PSF. \label{fig:psf} }
\end{figure}

\begin{figure} [ht]
\begin{center}
\begin{tabular}{c} 
\includegraphics[height=7cm]{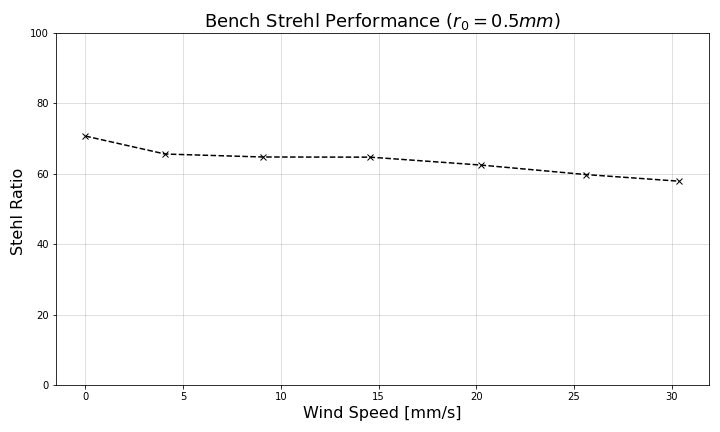}
\end{tabular}
\end{center}
\caption[example] 
{ The estimated Strehl ratio across a range of phase screen speeds. The laser source approximated a 3 mag J-band guide star. The SAPHIRA avalanche gain was set to unity and no modulation was used. The non-aberrated Strehl ratio of our testbed is about 85\% from alignment and optical errors. \label{fig:bench_perform} }
\end{figure}

In order to quantify the performance of our system, we compute the Strehl ratio relative to a best-fit airy model (see Figure~\ref{fig:psf}).
During closed-loop operations, photos of the corrected PSF were captured by a high speed infrared camera.
The Strehl ratio was estimated through a multi-step process: the images were stacked together to produce a `science' PSF approximating a long exposure; an airy disk model was assumed based on the best fit to the unabberrated testbed PSF; the empirical PSF was shifted to get a more accurate peak estimate; the Airy model was down-sampled and shifted to match the empirical PSF; both images were normalized and the Strehl ratio was estimated to be the ratio of their central values. This process was repeated as a function of phase screen speed, producing Figure~\ref{fig:bench_perform}. 
We observe a Strehl ratio of about 65\% with a slight degradation as wind speed increased.
This level of performance meets our expectation for our experimental set-up and turbulence parameters. 
This shows that the SAPHIRA assembly has met minimum requirements for integration into the MAPS WFS.

\section{FUTURE WORK}

\subsection{Status}

The NIR PyWFS has officially completed its commissioning phase and has entered its integration phase. 
In June 2022, the camera was shipped to the Steward Observatory at the University of Arizona where it is currently being integrated into the MAPS WFS unit.
Once the camera is installed into the WFS unit and its operation has been verified, we will begin the process of coordinating with the other WFS components and the MAPS control software.
The current timeline projects that the MAPS WFS and ASM will begin on-sky testing at MMTO in the Fall of 2022.

\subsection{Calibration}

Typically, AO systems operate as a self-contained instrument which is fed directly from the focal plane of a telescope. 
MAPS's adaptive secondary, however, is an example of a `pre-focal' deformable mirror with no intermediate focus. 
While adaptive secondary mirrors have several key advantages over their `post-focal' counterparts, this configuration substantially increases the complexity of calibration.
Systems with concave ASMs such as the Large Binocular Telescope (LBT) have overcome this issue by installing calibration sources at their intermediate focus; however, MMT uses a convex ASM with no access to a usable focal plane.
This is a well-known problem in the field for which several solutions have been theorized, investigated, and implemented.
Without access to a calibration source, there are two options to calibrate the system: use on-sky data; or generate a synthetic calibration matrix based on detailed simulations of the system. \cite{ASM_calib_overview}\cite{ASM_calib_first_high_order}\cite{ASM_CALIB_STRATEGY}
Synthetic calibration is a good option for systems which are only able to correct low spatial frequency aberrations since simulations tend to perform well at those spatial frequencies.
Since MMTO's former AO system was not sensitive to higher order aberrations ($\sim 50$ modes), it was able to use a synthetic calibration.
Synthetic calibration for higher order systems such as the new MAPS AO system have been previously implemented\cite{ASM_calib_first_high_order}, but require a substantial amount of empirical data about the adaptive secondary and wavefront sensor.
Obtaining enough empirical data to verify a high-order simulation is much more difficult for convex adaptive secondaries for the same reason as calibration.

Using on-sky data is non-trivial since the wavefront sensor will simultaneously measure both the atmospheric and induced aberrations, making it difficult to recover the ASM's influence on the wavefront error during calibration.
Pinna et al. 2012\cite{ASM_calib_first_high_order} reported on the first high order calibration of an adaptive optics system using on-sky data.
Their methodology employed a sinusoidal modulation technique whereby the magnitude a given basis ASM shape is varied over time with a known frequency.
After measuring several cycles of this modulated signal, the AO system telemetry will now have a distinct feature at the modulation frequency from which the calibration matrix can be determined.
The authors were able to compare system performance using a traditional calibration to both a synthetic and on-sky calibration.
They concluded that a combination of synthetic and on-sky methods is likely to produce the best results.
Later, hybrid methods such as the SPRINT calibration strategy \cite{SPRINT_ASM_hybrid_cal} defined feedback loops that adjust simulated calibration matrices using on-sky data, potentially eliminating the requirement for sophisticated measurements of the ASM.
We are currently investigating the implementation of a hybrid calibration strategy for MAPS, but may additionally implement purely empirical methods such as the recent DO-CRIME algorithm \cite{ASM_CALIB_DOCRIME}.

\subsection{Conclusion}

This proceedings reported on the status of the near-infrared pyramid wavefront (NIR PyWFS) sensor currently being commissioned as part of the MAPS upgrade to the MMT observatory. Advancements in avalanche photodiode arrays, such as Leonardo's SAPHIRA, have made NIR wavefront sensing a powerful tool for increasing NGS AO sky coverage. To date, readout electronics, firmware, and a cryostat Dewar have been developed for the wavefront sensor. Additionally, our SAPHIRA detector has undergone a suite of characterization testing and been operated in closed-loop on the AO testbed. Recently, the camera and cryostat assembly have been installed into the MAPS wavefront sensor unit at the University of Arizona with first light scheduled for the Fall of 2022. 



\acknowledgments 
 
The MAPS project is primarily funded through the NSF Mid-Scale Innovations Program, programs AST-1636647 and AST-1836008. The NIR PyWFS contributions have been funded by the Canada Foundation of Innovation and the Ontario Research Funds.

\bibliography{main} 
\bibliographystyle{spiebib} 

\end{document}